\documentclass[aps,twocolumn,amsmath,amsfonts,nofootinbib,preprintnumbers,
  superscriptaddress,eqsecnum,secnumarabic]{revtex4-1}
\usepackage{graphicx,slashed,hyperref}
\pdfoutput=1
\usepackage[utf8]{inputenc}
\hypersetup{
   bookmarks=true,         % show bookmarks bar?
   unicode=true,          % non-Latin characters in Acrobat�s bookmarks
   pdftoolbar=true,        % show Acrobat�s toolbar?
   pdfmenubar=true,        % show Acrobat�s menu?
   pdffitwindow=false,     % window fit to page when opened
   pdfstartview={FitH},    % fits the width of the page to the window
   pdftitle={My title},    % title
   pdfauthor={Author},     % author
   pdfsubject={Subject},   % subject of the document
   pdfcreator={Creator},   % creator of the document
   pdfproducer={Producer}, % producer of the document
   pdfkeywords={keyword1} {key2} {key3}, % list of keywords
   pdfnewwindow=true,      % links in new PDF window
   colorlinks=true,       % false: boxed links; true: colored links
   linkcolor=blue,          % color of internal links (change box color with linkbordercolor)
   citecolor=magenta,        % color of links to bibliography
   filecolor=magenta,      % color of file links
   urlcolor=blue           % color of external links
}
\usepackage{amsmath}
\usepackage{amssymb}
\usepackage{comment}
\newcommand{\be}{\begin{equation}}
\newcommand{\ee}{\end{equation}}
\def\bsp#1\esp{\begin{split}#1\end{split}}

\newcommand{\sidm}{{\sigma_{\rm IDM}}}
\begin{document}

\title{Dark Matter microphysics and 21 cm observations}

\author{Laura~Lopez-Honorez}
\email{llopezho@ulb.ac.be}
\affiliation{Service de Physique Th\'eorique, CP225, Universit\'e Libre de Bruxelles, Bld du Triomphe, 1050 Brussels, Belgium}
\affiliation{Vrije Universiteit Brussel and The International Solvay Institutes,
  Pleinlaan 2, 1050 Brussels, Belgium.}

\author{Olga Mena}
\email{olga.mena@ific.uv.es}
\author{Pablo Villanueva-Domingo}
\email{pablo.villanueva@ific.uv.es}
\affiliation{Instituto de F\'isica Corpuscular (IFIC), CSIC-Universitat de Valencia,\\ 
Apartado de Correos 22085,  E-46071, Spain}
\preprint{ULB-TH/18-13}
\begin{abstract}
  Dark matter interactions with massless or very light Standard Model
  particles, as photons or neutrinos, may lead to a suppression of the
  matter power spectrum at small scales and of the number of low mass
  haloes. Bounds on the dark matter scattering cross section with
  light degrees of freedom in such interacting dark matter (IDM)
  scenarios have been obtained from e.g. early time cosmic microwave
  background physics and large scale structure observations.  Here we
  scrutinize dark matter microphysics in light of the claimed $21$~cm
  EDGES 78 MHz absorption signal. IDM is expected to delay the 21 cm
  absorption features due to collisional damping effects. We identify
  the astrophysical conditions under which the existing constraints on the
    dark matter scattering cross section could be largely improved due to the IDM
    imprint on the 21 cm signal, providing also an explicit comparison to the WDM scenario.
\end{abstract}

\maketitle

% =====================================================================
\section{Introduction}

Interacting Dark Matter (IDM) with standard model light or massless
degrees of freedom, such as photons and neutrinos, gives rise to a
suppression of the small-scale matter power
spectrum~\cite{Boehm:2003xr,Boehm:2014vja, Schewtschenko:2014fca,
  Schewtschenko:2015rno} (see also Refs.~\cite{Cyr-Racine:2013fsa,
  Buckley:2014hja, Vogelsberger:2015gpr,
  Cyr-Racine:2015ihg,Aarssen:2012fx,Bringmann:2016ilk} for
interactions with new (dark) light degrees of freedom). This damping is
similar to the one caused by the free streaming of warm dark
matter (WDM). In IDM scenarios, in contrast, the suppression of the
small-scale overdensities is due to collisional
damping~\cite{Boehm:2000gq, Boehm:2001hm, Boehm:2004th}. These two
alternatives to the standard $\Lambda$CDM model fall into the category
of \emph{non-cold dark matter}
scenarios~\cite{Murgia:2017lwo} (NCDM). Such dark
matter models can provide some solutions to the the $\Lambda$CDM
 (where dark matter is made of purely cold and collisionless dark matter
particles) \textit{small scale crisis} (see e.g. the review of
Ref.~\cite{Bullock:2017xww}). A large number of studies in the
literature have been devoted to constrain the NCDM picture by means of
cosmological probes, such as Cosmic Microwave Background (CMB) fluctuations and spectral
distortions, galaxy clustering and Lyman-$\alpha$ forest power
spectrum, the number of Milky Way satellites, the reionization
history, or gravitational lensing ~\cite{Boehm:2003xr, Boehm:2004th,
  Viel:2013apy,Viel:2005qj, Seljak:2006qw, Viel:2006kd, Mangano:2006mp,
  Viel:2007mv, Boyarsky:2008xj,Serra:2009uu, McDermott:2010pa,
  Viel:2013apy, Wilkinson:2013kia, Dolgov:2013una, Wilkinson:2014ksa,
  Boehm:2014vja, Schneider:2014rda, Schewtschenko:2014fca,
  Escudero:2015yka, Ali-Haimoud:2015pwa, Baur:2015jsy,
  Schewtschenko:2015rno, Diamanti:2017xfo, Irsic:2017ixq,
  Yeche:2017upn, Gariazzo:2017pzb, Murgia:2017lwo, Diacoumis:2017hff,
  DiValentino:2017oaw, Campo:2017nwh,
  Stadler:2018jin,Rivero:2018bcd,Lovell:2017eec, Dayal:2014nva,
  Bose:2016hlz, Lopez-Honorez:2017csg,Barkana:2001gr, Yoshida:2003rm,
  Somerville:2003sh, Yue:2012na, Pacucci:2013jfa, Mesinger:2013nua,
  Schultz:2014eia, Dayal:2014nva, Lapi:2015zea,Ali-Haimoud:2015pwa,
  Bose:2016hlz, Bose:2016irl, Corasaniti:2016epp,Menci:2016eui,
  Lopez-Honorez:2017csg, Villanueva-Domingo:2017lae,
  Das:2017nub,Escudero:2018thh}.

In this regard, the 21 cm signal offers a new cosmological probe,
complementary to the existing ones, that could open a new window on
the early universe and can further test the imprint of NCDM (see
e.g.~\cite{Sitwell:2013fpa,Carucci:2015bra} for early work on the
subject). Here we will focus on the cosmic dawn period and in
particular on the first claimed detection of an absorption feature in
the sky-averaged global 21~cm signal at a redshift $z \sim 17$ by the
experiment to Detect the Global Epoch of Reionization Signatures
(EDGES)~\cite{Bowman:2018yin}.  The measured amplitude of the dip in
the 21~cm global signal appears to be much deeper than that expected
in standard CDM scenarios and therefore requires new physics to heat
the radio background or cool the gas temperature, see
also~\cite{Hills:2018vyr,Bradley:2018eev} for other possible
interpretations. This has triggered a surge of interest from the dark
matter community trying to relate this effect to dark matter decay and
annihilation~\cite{Berlin:2018sjs,DAmico:2018sxd,Liu:2018uzy,Mitridate:2018iag}~\footnote{
  See also the previous works of Refs.~\cite{Valdes:2012zv,
    Evoli:2014pva,Lopez-Honorez:2016sur}.}, and investigating the dark
matter scenarios that could account for the signal, see
e.g.~\cite{Pospelov:2018kdh,
  AristizabalSierra:2018emu,Boddy:2018wzy,Kovetz:2018zan,Barkana:2018qrx,Barkana:2018lgd}.

While NCDM scenarios are unlikely to explain the large absorption
amplitude, they delay structure formation and, therefore, might delay
the onset of reionization and of UV and X-ray emission, see
e.g.~\cite{Sitwell:2013fpa,Bose:2016hlz,Lopez-Honorez:2017csg,Escudero:2018thh}. As
a result a shift to later times in the typical features in the 21~cm
sky-averaged global signal and power spectrum is observed in the
context of non-cold dark matter~\cite{Sitwell:2013fpa,
  Escudero:2018thh,Schneider:2018xba,Safarzadeh:2018hhg,Lidz:2018fqo}.
Consequently, it is timely to study the compatibility between the
observation reported by EDGES, located at a redshift around $z\simeq
17$ and the IDM scenario. We follow two possible avenues. The first of
them relies on exploiting the non-negligible Lyman-$\alpha$ coupling
between the gas and the spin temperature characterizing the 21 cm
signal at $z\simeq 20$~\cite{Safarzadeh:2018hhg,Lidz:2018fqo}. The second one consists
in imposing the minimum in the absorption feature to happen before
$z\simeq 17$~\cite{Schneider:2018xba}. For both strategies, a number
of degeneracies between the details of dark matter microphysics and
the astrophysical parameters will appear. We will briefly discuss
their impact on the constraints on NCDM scenarios.

Universal fits to the halo mass functions from N-body
simulations for the IDM scenarios have been obtained
in~\cite{Schewtschenko:2014fca,Moline:2016fdo,Schewtschenko:2016fhl}. In
particular, here we will use the results of~\cite{Moline:2016fdo}
derived for IDM scenarios involving dark matter-photon scatterings. 
A possible particle physics model related to this cosmological
scenario is the case of millicharged dark
matter~\cite{Dvorkin:2013cea,Vinyoles:2015khy}. 
IDM scenarios including dark matter-neutrino scatterings have been shown to give rise to a very
similar damping in the power spectrum and also to a very similar mass
function as for dark matter-photon scatterings, see
Refs.~\cite{Boehm:2001hm,Schewtschenko:2014fca} and the Appendix \ref{sec:hmf}. 
Unfortunately, in the latter case, no publicly available dedicated analysis provides the necessary fits to
the associated halo mass functions necessary for our study. We
therefore use the IDM scattering on photons as a toy model to evaluate the
impact of the EDGES signal on the more general case of IDM with light
degrees of freedom. In order to ease comparison with previous studies,
we shall also study the case of thermal warm dark matter (WDM) with
mass in the keV range (see Refs.~\cite{Bode:2000gq, Knebe:2001kb,
  Colin:2007bk, Zavala:2009ms, Smith:2011ev, Lovell:2011rd,
  Schneider:2011yu, Polisensky:2013ppa, Lovell:2013ola,
  Kennedy:2013uta, Destri:2013hha, Angulo:2013sza, Benson:2012su,
  Kamada:2013sh, Lovell:2015psz, Ludlow:2016ifl, Wang:2016rio,
  Lovell:2016nkp, Nakama:2017ohe,Barkana:2001gr, Yoshida:2003rm,
  Somerville:2003sh, Yue:2012na, Pacucci:2013jfa, Mesinger:2013nua,
  Schultz:2014eia, Dayal:2014nva, Lapi:2015zea, Bose:2016hlz,
  Bose:2016irl, Corasaniti:2016epp,Menci:2016eui,
  Lopez-Honorez:2017csg, Villanueva-Domingo:2017lae,
  Das:2017nub,Irsic:2017ixq, Yeche:2017upn} and the most recent works
of Refs.~\cite{Schneider:2018xba,Lidz:2018fqo}).

The structure of the paper is as follows. We start in Sec.~\ref{sec:21cm}
by describing the physics of the 21~cm global signature. We account
for the effect of IDM in the 21~cm global signature in
Sec.~\ref{sec:cdm}, presenting the constraints on the dark matter
photon elastic cross sections arising from \textit{(i)} the presence
of a rich Lyman-$\alpha$ background at $z\simeq 20$ (see
Sec.~\ref{sec:c2}), and \textit{(ii)} the location of the
EDGES minimum (see Sec.~\ref{sec:c1}). Finally, we summarize our results and conclude in Sec.~\ref{sec:concl}.

%%%%%%%%%%%%%%%%%%%%
\section{The 21 cm signal}
\label{sec:21cm}
%%%%%%%%%%%%%%%%%%%

%%%%%%%%%%%%%%%%%%%%
\subsection{The differential brightness temperature}
\label{sec:dTb}
%%%%%%%%%%%%%%%%%%%

The brightness of a patch of neutral hydrogen (HI) relative
to the CMB at a given redshift $z$ is expressed in terms of the differential brightness
temperature, $\delta T_b$. The sky-averaged $\delta T_b$ scales
as~\cite{Madau:1996cs, Furlanetto:2006jb, Pritchard:2011xb,
  Furlanetto:2015apc}
\begin{widetext}
\begin{align}
\frac{\delta T_b(\nu)}{\textrm{mK} } &\simeq 27  x_\textrm{HI}   \left( 1 - \frac{T_\textrm{CMB}}{T_{\rm S}}\right) \left( \frac{1+z}{10}\right)^{1/2} \left(\frac{0.15}{\Omega_m h^2} \right)^{1/2} \left( \frac{\Omega_b h^2}{0.023}\right) ~, 
\label{eq:Tbdev}
\end{align}
\end{widetext}
where $\nu= \nu_{21}/(1+z)$ with $\nu_{21}= $ 1420 MHz,  $x_\textrm{HI}$ represents the fraction of neutral hydrogen and
$\Omega_b h^2$ and $\Omega_m h^2$ refer to the current baryon and
matter contributions to the universe's mass-energy content. The ratio
of the populations of the two ground state hyperfine levels of
hydrogen is quantified by the spin temperature, $T_{\rm S}$, which is
determined by three competing effects~\cite{Hirata:2005mz}: 1)
absorption and stimulated emission of CMB photons coupling the spin
temperature to the CMB temperature $T_{\rm CMB}$ in contrast with 2)
atomic collisions (which are important at high redshifts $z\gtrsim
30$); and 3) resonant scattering of Lyman-$\alpha$ photons that couple
the spin temperature to the gas kinetic temperature $T_k$. The latter
effect is the so-called Wouthuysen-Field effect~\cite{Wouthuysen:1952,
  Field:1958} that turns on with the first sources. Assuming that the Lyman-$\alpha$ color temperature is
$T_\alpha \simeq T_k$~\cite{Pritchard:2011xb}, the spin
temperature results from:
\begin{equation}
  \left( 1 - \frac{T_\textrm{CMB}}{T_{\rm S}}\right)=\frac{x_{tot}}{1+x_{tot}} \left( 1 - \frac{T_\textrm{CMB}}{T_k}\right)
\label{eq:TgamTsappr}
\end{equation}
with $x_{tot}= x_\alpha+ x_c$ and $x_c$ and $x_\alpha$ are the
coupling coefficients for collisions and Lyman-$\alpha$
scatterings.

At the low redshifts of interest here, collision coupling
effects can be safely neglected and therefore, $x_{tot}=x_\alpha$. The
Lyman-$\alpha$ coupling is defined as
\begin{equation}
x_\alpha = \frac{16\pi^2T_{\star}e^2f_{\alpha}}{27A_{10}T_{\gamma}m_e c} S_{\alpha}J_{\alpha},
\label{x_alpha}
\end{equation}
where $T_\star = h \nu_{21}=$ 68.2 mK is the hyperfine energy
splitting, $e$ and $m_e$ the charge and mass of the electron, $
f_{\alpha}$ is the oscillator strength of the Lyman-$\alpha$
transition, $A_{10}$ is the spontaneous decay rate of the 21 cm
transition, $J_\alpha$ is the specific intensity of the background
radiation evaluated at the Lyman-$\alpha$ frequency and $S_\alpha$ is
an order unity correction factor which accounts for the detailed shape
of the spectrum near the resonance \cite{Chen:2003gc}. In particular,
in the framework considered here, $S_\alpha \lesssim 1$ and the
Lyman-$\alpha$ flux gets two types of contributions. One results from
the X-ray excitation of HI ($J_{\alpha X}$), while the other one
results from direct stellar emission of UV photons between
Lyman-$\alpha$ and the Lyman-$\alpha$ limit $J_{\alpha \star}$, thus
$J_\alpha=J_{\alpha X}+ J_{\alpha \star}$~\cite{Mesinger:2010ne}. We
will see in Sec.~\ref{sec:dTb} that for the X-ray efficiencies
considered here, $J_{\alpha X}$ only represents a small contribution
to the total Ly$\alpha$ flux. On the other hand, the direct stellar
emission contribution to $J_{\alpha}$ is computed assuming by default
a Pop II stars spectral model. This gives rise to the emission of 9690
photons per baryon between Lyman-$\alpha$ and the Lyman limit, see
Appendix~\ref{sec:appb} (see also ~\cite{Barkana:2004vb}).  Notice
from Eq.~(\ref{eq:TgamTsappr}) that, when $x_{tot}=x_\alpha=1$,
$\delta T_b$ will be at the half of the value that it would have if
$T_S$ were completely coupled to $T_k$, which happens when $x_\alpha
\gg 1$.  The authors of Ref.~\cite{Lidz:2018fqo}, following the EDGES
results~\cite{Bowman:2018yin}, have imposed that $x_\alpha$ should be
one or larger at redshift $z\simeq 20$. We shall apply this constraint
in our numerical analyses of IDM scenarios, see Sec.~\ref{sec:c2}.

\begin{figure}[t]
\includegraphics[width=.5\textwidth]{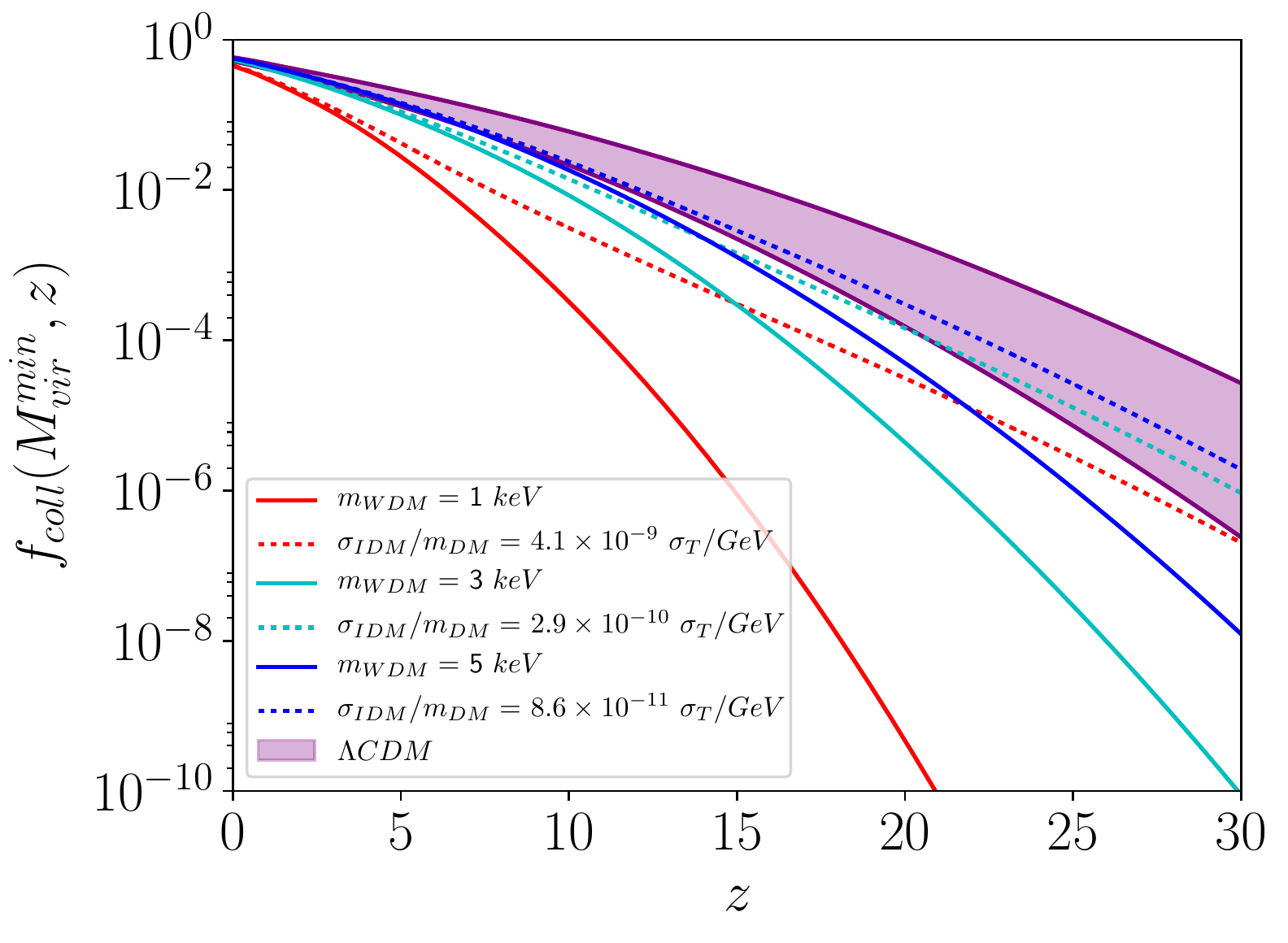}
\caption{Fraction of mass collapsed into haloes of mass larger than
  $M_{\rm vir}^{\rm min} (z=20) \simeq 10^6 \; M_{\odot}$
  (corresponding to $T_{\rm vir}^{\rm min}=10^3$ K) as a function of
  the redshift.  The continuous lines depict the WDM cases that are
  compared with IDM scenarios, shown with dotted curves. Continuous
  and dotted curves of the same colors corresponds to a fixed value of
  the half-mode mass. The purple region illustrates the change in the
  CDM collapsed fraction (using the Sheth-Tormen mass function) for
  $T_{\rm vir}^{\rm min}$ varying within the range of $10^3$ to $10^4$
  K.}
\label{fig:fcoll}
\end{figure}

In order to extract the imprint of NCDM on the 21~cm signal, we profit
from the publicly available tool {\tt 21cmFast}. The main
purpose of the code is the study of variations in the 21~cm signal due
to changes in a given set of astrophysical and cosmological
parameters. We make use of the output values of $x_\alpha$ and of $T_S$ and $x_{\rm HI}$
to extract the differential brightness temperature as in
Eq.~(\ref{eq:Tbdev}). In practice, we use a version of the {\tt
  21cmFast} code adapted to account for the IDM and the
WDM as detailed in Ref.~\cite{Escudero:2018thh}.  We have modified the
default WDM scenario implementation modifying the definition of both
the transfer function and the halo mass function according to the
prescription given in the Appendix~\ref{sec:hmf}.  This halo mass function plays
an important role in the evaluation of the production rate of
ultraviolet (UV), X-rays, and Lyman-$\alpha$ radiation, responsible of
the ionization, heating and Lyman-$\alpha$ coupling
respectively. These production rates are proportional to the star
formation rate $\dot{\rho}_{\star}$. In {\tt 21cmFast}, this quantity
is evaluated in terms of the growth of the fraction of mass collapsed
in haloes which are able to host star-forming galaxies, $f_{\rm
  coll}(>M_{\rm vir}^{\rm min})$, defined as
\begin{equation}
f_{ \rm coll} (>M_{\rm vir}^{\rm min}) = \frac{1}{\rho_{m,0}} \int_{M_{\rm vir}^{\rm min}}^{\infty} \, M \, \frac{dn}{dM} \, dM,
\label{eq:fcoll}
\end{equation}
where $\rho_{m,0}$ is the current matter density and ${dn}/{dM}$ is
the halo mass function. For the NCDM cosmologies, the halo mass
function is always suppressed at small masses compared to the CDM
scenario, giving rise to a smaller $f_{\rm coll}$, at fixed redshift,
as illustrated in Fig.~\ref{fig:fcoll}, see
also~\cite{Sitwell:2013fpa, deSouza:2013hsj, Boehm:2003xr}.\footnote{Let us emphasize that our prescription for WDM differs from
  one of~\cite{Sitwell:2013fpa}. Here we follow the fits to the
  results of simulations from~\cite{Schneider:2011yu} for WDM which
  was directly compared to the results of IDM simulations
  in~\cite{Schewtschenko:2014fca,Moline:2016pbm,Schewtschenko:2016fhl}.
  Reference~\cite{Sitwell:2013fpa} followed an earlier prescription
  introduced by~\cite{Barkana:2001gr}.} With the
purple area we show the case of CDM with threshold masses $M_{\rm
  vir}^{\rm min}$ between $10^6 M_\odot$ (upper curve) and
$3\times10^7M_\odot$ (lower curve) at $z=20$ or equivalently $T_{\rm
  vir}^{\rm min}$ between $10^3$  and $10^4$ K, see
eq.~(\ref{eq:mminT}). The upper purple curve for CDM can be compared
to case of NCDM scenarios in the form of WDM (red, blue and cyan
continuous curves) and of IDM (red, blue and cyan dotted curves) for
the same $M_{\rm vir}^{\rm min}=10^6 M_\odot$ at $z=20$ ($T_{\rm
  vir}^{\rm min} = 10^3$ K).  The IDM scattering cross-sections are
normalized in terms of the Thompson cross section $\sigma_T = 6.65
\times 10^{-25}$~cm$^2$.

%% Even though the formation of haloes
%capable
%% of star formation is delayed in NCDM, the rate at which it increases
%% with redshift is thus faster than in CDM as already noted
%% in~\cite{Sitwell:2013fpa}, see also~\cite{deSouza:2013hsj,
%%   Boehm:2003xr}.

Overall, in the framework considered here, the NCDM modification of
$f_{\rm coll}$ will result into a delayed reionization, heating and
Lyman-$\alpha$ coupling, giving rise to an absorption feature in the
21 cm signal located at lower redshifts with respect to CDM
scenarios~\cite{Sitwell:2013fpa,Bose:2016hlz,Lopez-Honorez:2017csg,Escudero:2018thh}.
Notice also that curves of a fixed color in Fig.~\ref{fig:fcoll}
correspond to IDM and WDM models giving rise to a fixed value of the
half-mode mass or breaking scale of the linear power spectrum, see
Eqs.~(\ref{eq:alphWDM}), (\ref{eq:alphIDM}) and (\ref{eq:Mhm}) in
Appendix A. For fixed half-mode mass, the suppression of the $f_{\rm
  coll}$ is always more severe in the WDM scenarios than in the IDM
scenarios due to a relatively larger number of low mass haloes in the
IDM case, see~\cite{Moline:2016fdo}. The rate at which $f_{\rm coll}$
increases with redshift is also different between IDM and WDM. These
features might help to discriminate between NCDM models using both the
21 cm global signal and its power spectrum~\cite{Escudero:2018thh}.

%

%%%%%%%%%%%%%%%%%%%%
\subsection{The astrophysical parameters}
\label{sec:dTb}
%%%%%%%%%%%%%%%%%%%
The minimum virial mass $M_{\rm vir}^{\rm
  min}$ from which haloes begin to efficiently form stars (see  Eq.~(\ref{eq:fcoll})) is related
to the threshold temperature $T_{\rm vir}^{\rm min}$ as~\cite{Barkana:2000fd}
 \begin{equation}
\label{eq:mminT}
M_{\rm vir}^{\rm min} (z) = 10^8 \left(\frac{T_{\rm vir}^{\rm min}}{1.98 \times 10^4 \, {\rm K}} \frac{0.6}{\mu} \right)^{3/2} \left(\frac{1+z}{10}\right)^{-3/2} M_\odot/h ~,
\end{equation}
  where $\mu$ is the mean molecular weight and it is equal to 1.2 (0.6) for a
  neutral (fully ionized) primordial gas.
  The value of the minimum virial temperature depends on the cooling
  mechanism considered. The lower threshold to make the atomic cooling
  channel effective is $T_{\rm vir}^{\rm min}=
  10^4$~K~\cite{Evrard:1990fu, Blanchard:1992,Tegmark:1996yt,
    Haiman:1999mn, Ciardi:1999mx}, corresponding to $M_{\rm vir}^{\rm
    min}= 3\times 10^7 M_\odot$ at $z=20$ for $\mu=0.6$. In contrast,
  the molecular $H_2$ cooling channel can be effective down to
  temperatures of $T_{\rm vir}^{\rm min}= 10^3$~K corresponding to $M_{\rm vir}^{\rm
    min}= 10^6 M_\odot$ at $z=20$. The hydrogen molecules
  may however be destroyed by the ionizing radiation. Nevertheless, several
  hydrodynamical works in the
  literature~\cite{Wise:2007cf,Wise:2014vwa} have shown that in the
  presence of a large soft UV background, molecular cooling could be
  highly effective, i.e. metal-enriched star formation is not
  restricted to atomic cooling. Furthermore, even in the absence of
  electrons, molecular cooling could cool down the gas in haloes
  associated to virial temperatures much lower than the ones required
  for atomic cooling. Consequently, in the following, 
  we shall consider $T_{\rm vir}^{\rm min}= 10^3$~K as the minimum
  threshold temperature for star formation and we assume the same
  threshold temperature $T_{\rm vir}^{\rm min}$ for haloes hosting
  ionizing and X-ray sources. This parameter plays a crucial 
  role in extracting the constraints from the 21 cm
  absorption signal in both CDM and NCDM scenarios, see
  also~\cite{Schneider:2018xba,Safarzadeh:2018hhg}. The impact of
  varying $T_{\rm vir}^{\rm min}$ in the $10^3$~K to
  $10^4$~K range within the CDM paradigm is illustrated in Fig.~\ref{fig:fcoll} with the
  purple area. The lower value of the threshold parameter corresponds to
  higher values of the faction of collapsed haloes at a given
  redshift.

The comoving star formation rate density is described by
$\dot{\rho}_{\star} = f_{*}\rho_{b,0}\dot{f}_{ coll} (>M_{\rm
  vir}^{\rm min})$, where $f_*$, the fraction of baryons that are
converted into stars, considered here as a constant parameter (neglecting
any dependence on the halo masses or redshift). As $f_*$ controls
the amplitude of the star formation rate density, it also sets the
amplitude of both the ionizing and heating rates, as well as the
Lyman-$\alpha$ flux. This parameter is quite uncertain as no
  observations of low mass mass haloes of mass $10^6-10^8 \, M_\odot$ at
  redshift $z\sim 20$ are available. Nevertheless, several previous
works, based on radiation-hydrodynamic simulations of high-redshift
galaxies in a neutral medium~\cite{Wise:2014vwa, 2016ApJ...833...84X,
  Ma:2017avo, 2018MNRAS.479..994R} or based on the comparison of the star
formation rate density to the one derived from UV luminosity function
measurements~\cite{Lidz:2018fqo, 2016MNRAS.455.2101M,Leite:2017ekt}, 
 have found values of $f_*\sim {\cal O} (0.01)$. In the following, we
 therefore consider $f_*=0.01$, leaving for the
  discussion in Sec.~\ref{sec:concl} the impact in our results of
  slightly larger values of $f_*$.

Finally, in order to characterize the overall normalization of the
X-ray luminosity, {\tt 21cmFast} makes use of the X-ray efficiency
parameter $\zeta_X$, expressed in units of $M_{\odot}^{-1}$. This
parameter is varied here within the range $\zeta_X=[1-5]\times
10^{56}/M_{\odot}$. One can relate this parameter to the integrated
X-ray soft band emissivity (below 2 keV) per unit of star formation
rate escaping the galaxy $L_{X<2{\rm keV}}/{\rm SFR}$, varied in the range $ \log_{10}(L_{X<2{\rm keV}}/{\rm SFR})\in
[39.5,40.2]$ erg/s / ($M_{\odot}$/yr).~\footnote{ The emissivity
  $L_{X<2{\rm keV}}/{ \rm SFR}$ corresponds to the following
  combination of parameters: $\alpha_X\zeta_X h_p \int
  (\nu/\nu_0)^{-\alpha_X} d\nu$ where $h_p$ is the Planck constant,
  $\alpha_X$ is the spectral slope parameter and $\nu_0$ is the
  obscuration frequency cutoff parameter (see the {\tt 21CMMC}
  code~\cite{Greig:2015qca} based on {\tt 21cmFast}) . The integral
  goes from $\nu_0$ to 2 keV. We took $\alpha_X=1.2$ and $\nu_0= 7.2
  \times 10^{12}$ Hz (corresponding to an energy of 300 eV), that are
  the default values in the {\tt 21cmFast} code.}
This range is similar to one extracted from observations of the hot
interstellar medium, which lead to $\log_{10}(L_{X<2{\rm keV}}/{\rm
  SFR})\sim [39,40]$ erg/s / ($M_{\odot}$/yr)~\cite{Mineo:2012mq}, and
also to the one adopted in Ref.~\cite{Schneider:2018xba} (see also
Ref.~\cite{Monsalve:2018fno}).

\section{Imprint of the IDM on the 21 cm signal}
\label{sec:cdm}
\begin{figure}[t]
\includegraphics[width=.5\textwidth]{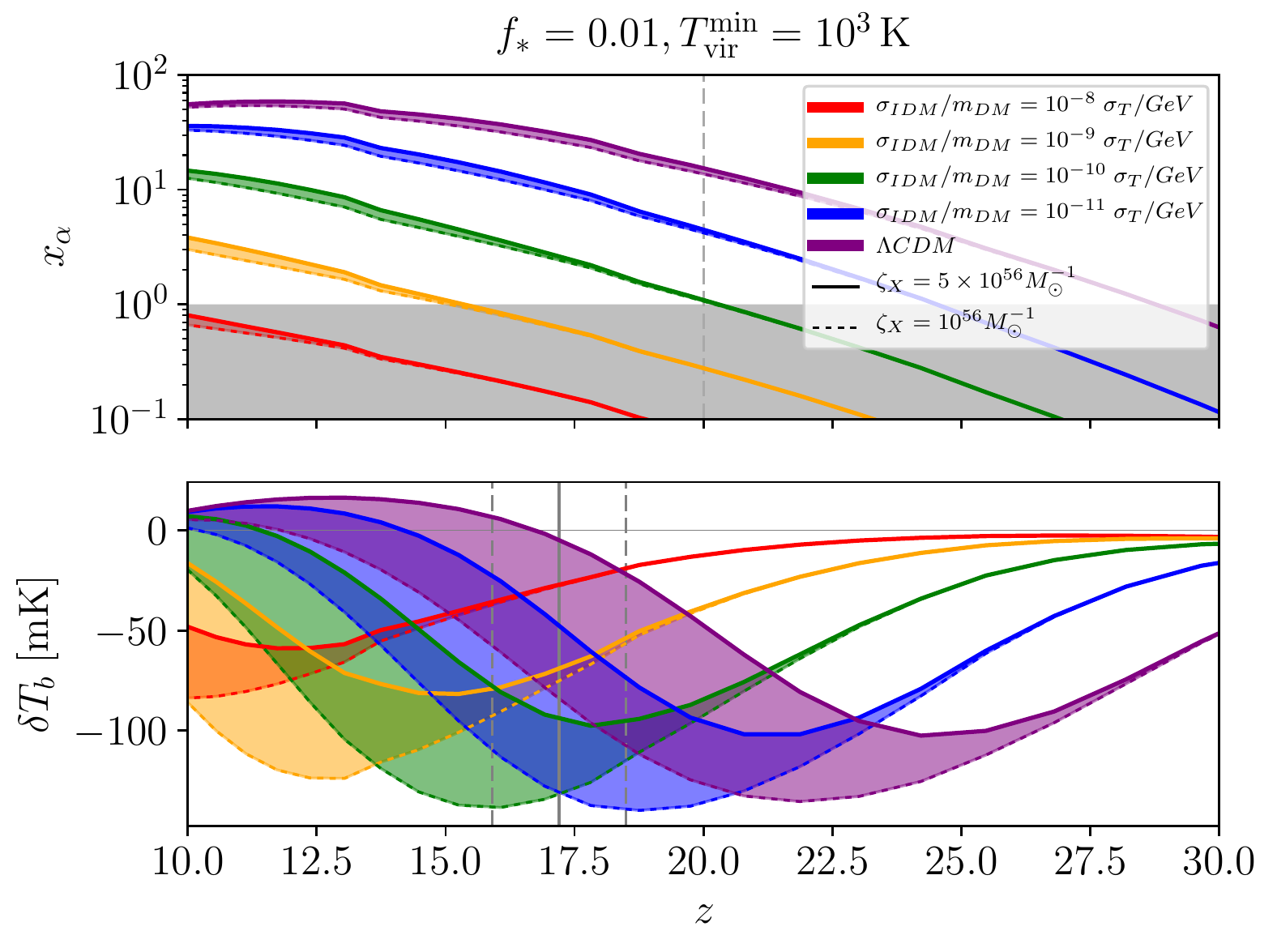}
\caption{Coupling coefficient for Lyman-$\alpha$ scattering (top) and sky-averaged 21~cm
  brightness temperature (bottom) as a function of the redshift for
  several possible values of the scattering dark matter-photon
  cross-section over the dark matter particle mass. We have fixed
 $f_*=0.01$ and $T_{\rm vir}^{\rm min}=10^3$~K. The width of
 the bands refers to the change in $\delta T_b$ and $x_\alpha$ due to
 different values of $\zeta_X$.}
\label{fig:Tbxa}
\end{figure}

 Non-cold dark matter scenarios are expected to delay structure
 formation and therefore the absorption feature in the 21~cm signal at
 cosmic dawn. This effect is illustrated for the IDM model under study
 in Fig.~\ref{fig:Tbxa}, where we show the coupling coefficient for
 Lyman-$\alpha$ scattering $x_\alpha$ (top), and the sky-averaged
 21~cm brightness temperature (bottom). The different colors
 correspond to different values of the dark matter-photon scattering
 cross-sections over the dark matter particle mass,
 $\sidm/\sigma_T({\rm GeV}/{m_{\rm DM}})$. 
As can be noticed, large cross sections, inducing a stronger suppression at small
 scales,  also induce a stronger suppression of the Lyman-$\alpha$
 coupling at a given redshift (see the top panel of Fig.~\ref{fig:Tbxa}) and a
 larger shift towards smaller redshifts of the 21~cm features (see the
 bottom panel of Fig.~\ref{fig:Tbxa}).  We also illustrate, with a 
 vertical line, the position of the minimum of absorption reported by
 the EDGES experiment (at $\nu=78$~MHz corresponding to $z=17.2$). The
 dashed lines correspond to the largest signal redshift range at the
 minimum of absorption within the 99\% CL interval reported by
 Ref.~\cite{Bowman:2018yin}.

  Figure~\ref{fig:Tbxa} also shows the impact of varying the X-ray
  heating efficiency $\zeta_X$ in the $10^{56} M_\odot^{-1}-5 \times
  10^{56} M_\odot^{-1}$ range with the width of the colored bands.  We
  see from the top panel of Fig.~\ref{fig:Tbxa} that the range of
  $\zeta_{X}$ considered here, the Lyman-$\alpha$ flux resulting from
  X-ray excitation ($J_{\alpha X}\propto\zeta_X$) is usually a negligible
  contribution to the Lyman-$\alpha$ coupling at the redshift at which
  $x_\alpha\sim 1$. On the bottom panel of Fig.~\ref{fig:Tbxa}, we see
  the impact of the X-ray heating parameter $\zeta_X$ on the
  differential brightness temperature. Here the impact is more
  significant. The deepest absorption dip corresponds to $\zeta_{X}=
  10^{56} M_\odot^{-1}$, while the shallowest one is obtained for
  $\zeta_{X}= 5 \times 10^{56} M_\odot^{-1}$.  Notice also that a
  larger value of $\zeta_{X}$ implies an earlier minimum, i.e. the
  X-ray heating of the IGM occurs earlier in time. As a result, in
  order to extract conservative constraints on the NCDM parameters
  from the redshift at which the absorption minimum is located, we
  shall consider the value $\zeta_{X}= 5\times 10^{56}/M_\odot$, see
  Sec.~\ref{sec:c1}.

Figure~\ref{fig:xa20} illustrates the equivalent to
Fig.~\ref{fig:Tbxa} with $\zeta_X$ fixed to $ 5 \times 10^{56}
M_\odot^{-1}$ and considering this time $T_{\rm vir}^{\rm min}= 10^3$
K (continuous curves) as well as $T_{\rm vir}^{\rm min}= 10^4$ K
(dotted curves). Going from left to right panel of Fig.~\ref{fig:xa20}
we increase the value of $f_*$ by a factor of 3. Based on these plots,
we discuss in the next two subsections the constraints that could be
derived on NCDM scenarios following two different approaches.

\subsection{ Constraints from the Lyman-$\alpha$ background}
\label{sec:c2}

\begin{figure*}[t]
  \hspace*{-1cm}
\includegraphics[width=.49\textwidth]{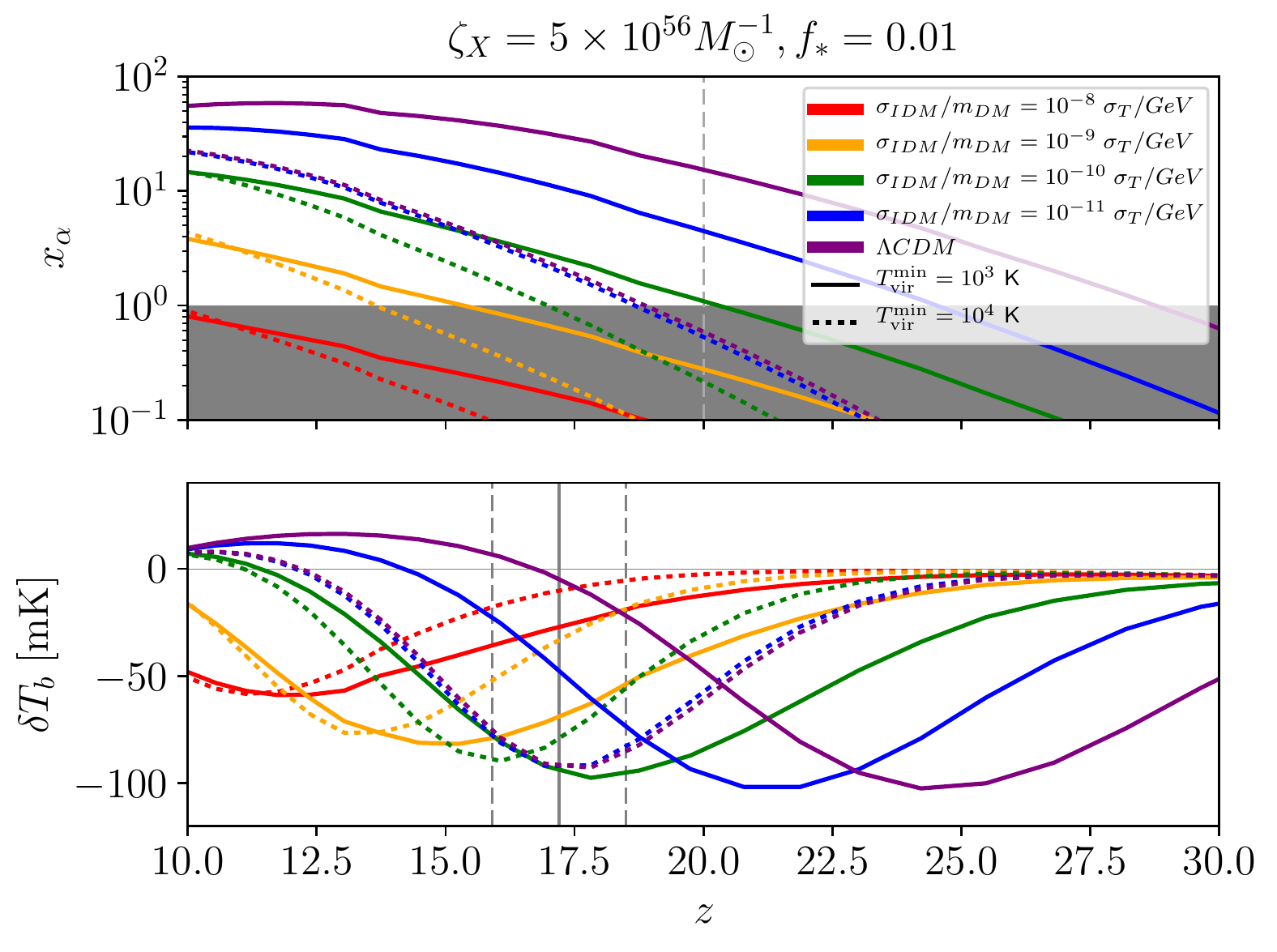}
\includegraphics[width=.49\textwidth]{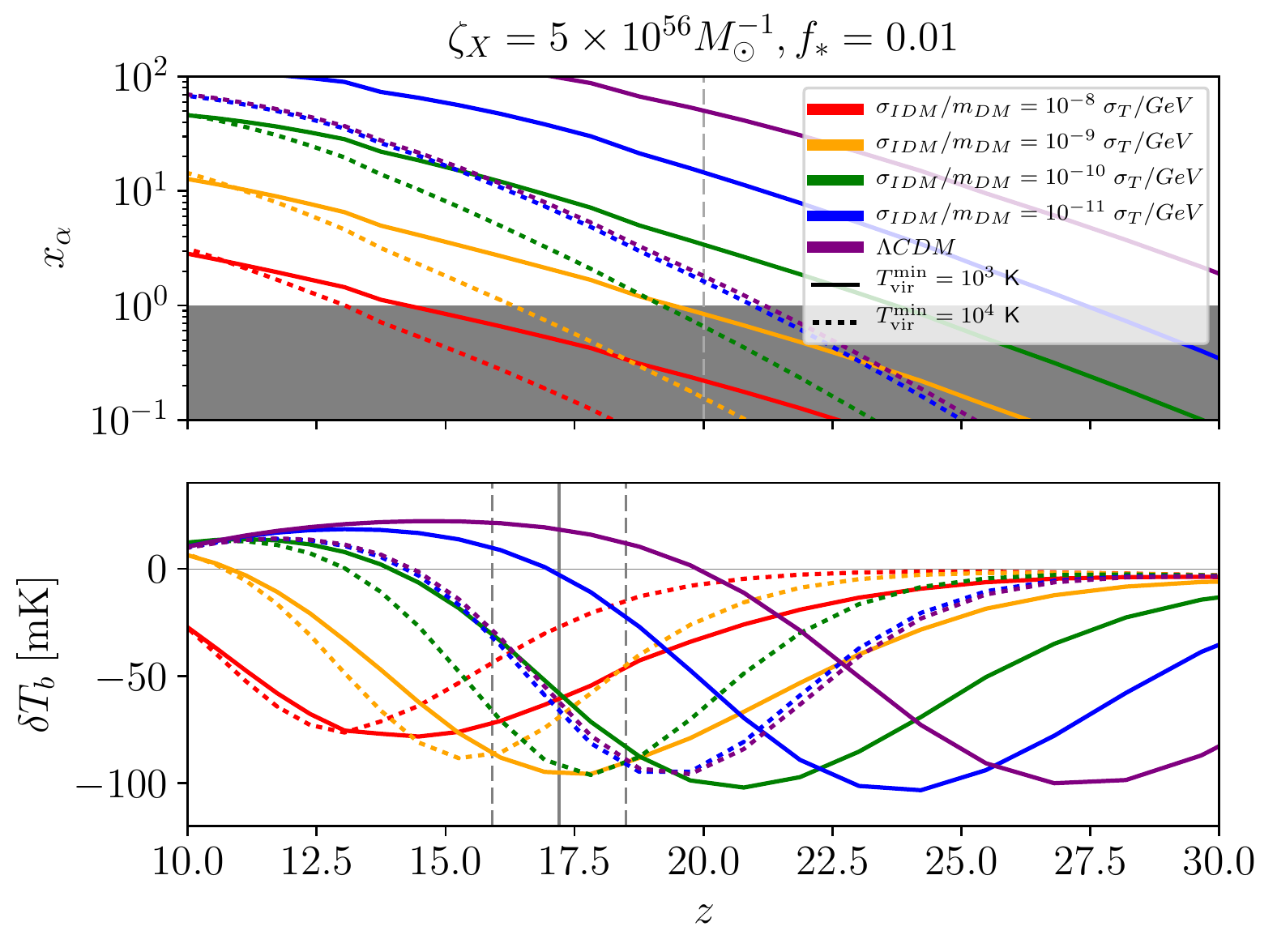}
\caption{Top panels: Lyman-$\alpha$ coupling coefficient $x_\alpha$
  versus redshift for different values of the astrophysical parameter
  $T_{\rm vir}^{\rm min}$  and for several possible values of the
  scattering dark matter-photon cross-section over the dark matter
  mass. Cases within the shaded region
  are highly disfavoured by the condition given by
  Eq.~(\ref{eq:xa20}). Bottom panels: sky-averaged 21~cm  brightness
  temperature. The solid (dashed) lines indicates the mean redshift
  (range in redshift) associated to the EDGES signal. We have fixed
  $\zeta_X=5 \times 10^{56}M_\odot$ 
and $f_*=0.01$ ($f_*=0.03$) in left (right) panel.}
\label{fig:xa20}
\end{figure*}

In order to account for the EDGES results, the authors of
Ref.~\cite{Lidz:2018fqo} based their analyses on the assumption that
a sufficiently strong Lyman-$\alpha$ background is present by $z \sim$
20. They  imposed
\begin{equation}
  x_\alpha (z=20)\gtrsim 1~.
\label{eq:xa20}
\end{equation}
This limit results from the observation that the absorption signal
reported by the EDGES experiment is equal to half of the maximum
amplitude of absorption at $z\simeq 20$.  We shall follow this
assumption here, applying this condition to our simulations within
NCDM models.  Let us first focus on the top left panel of
Fig.~\ref{fig:xa20}, which shows the Lyman-$\alpha$ coupling
coefficient $x_\alpha$ as a function of the redshift for
$f_*=0.01$. The condition reported in (\ref{eq:xa20}) disfavors the
cosmological scenarios associated to a prediction of $x_\alpha (z)$
lying within the shaded area at $z=20$.

  %% \footnote{ More precisely, for
  %% the best-fit signal, $x_\alpha (z_{1/2})= 1$ is obtained for
  %% $z_{1/2}=19.75$, the redshift at which $\delta T_b$ reaches half of
  %% the maximum amplitude of absorption. Varying the $\delta T_b$ within
  %% the 99\% CL range reported by EDGES, one can extract $z_{1/2}\in
  %% [19.45, 20.35]$. As can be noticed from Figs.~\ref{fig:Tbxa} and
  %% \ref{fig:xa20} imposing $x_\alpha (z_{1/2})= 1$ within this range
  %% will lead to very similar results.}

The astrophysical parameter $T_{\rm vir}^{\rm
  min}$ has a significant impact on the Lyman-$\alpha$ coupling
coefficient $x_\alpha$. Larger virial temperature shifts star
formation to lower redshifts, giving rise to a lower
Lyman-$\alpha$ background at a given redshift.  In a similar
way, a large value of the IDM scattering cross-section implies a longer delay in
structure formation. As a result, one can deduce from
Fig.~\ref{fig:xa20} (left panel) that, for molecular cooling ($T_{\rm vir}^{\rm
  min}=10^3$~K), the scattering cross-section must be below $10^{-10}
\, \sigma_T \times \left( {m_{\rm DM}}/{\rm GeV}\right)$.  If one
assumes instead that the only efficient cooling mechanism is atomic
cooling with $T_{\rm vir}^{\rm min} \geq 10^4$~K, the limits on the IDM
scattering cross section become much tighter, excluding scattering
cross-section lower than $10^{-11} \, \sigma_T \times \left( {m_{\rm
    DM}}/{\rm GeV}\right)$. Our most conservative bound (assuming
$f_*=0.01$) is therefore $\sidm
\lesssim 10^{-10} \, \sigma_T \times \left( {m_{\rm DM}}/{\rm
  GeV}\right)$. This constraint is stronger than the $95\%$~CL upper limit of
$\sidm< 8 \times 10^{-10} \, \sigma_T \times \left(
         {m_{\rm DM}}/{\rm GeV}\right)$ reported in
       Ref.~\cite{Escudero:2018thh}, 
based on observations of Milky Way satellite galaxy number counts and
assuming a mass for our galaxy of  $M_{\rm{MW}} = 2.6\times 10^{12}M_\odot$.

         Notice that the Lyman-$\alpha$ coupling coefficient
         $x_\alpha$ is directly proportional to the fraction of
         baryons converted into stars ($f_*$), here considered as
         constant.  The left (right) panels of Fig.~\ref{fig:xa20}
         have been simulated with $f_*=0.01$ ($f_*=0.03$). Notice that
         larger values of $f_*$ increase the Lyman-$\alpha$ coupling
         at a fixed redshift, weakening the bound on IDM scenarios
         resulting from Eq.~(\ref{eq:xa20}).  In the case of $f_*=0.03$, the limits quoted
         above are translated into $\sidm \lesssim
         10^{-9} \, \sigma_T \times \left( {m_{\rm DM}}/{\rm
           GeV}\right)$ and $\sidm \lesssim
         10^{-10} \, \sigma_T \times \left( {m_{\rm DM}}/{\rm
           GeV}\right)$ for $T_{\rm vir}^{\rm min}=10^3$~K and $T_{\rm
           vir}^{\rm min}=10^4$~K respectively.

 \subsection{ Constraints from the position of the absorption
   minimum in the $21$~cm global signature}
 \label{sec:c1}

Another possible avenue to constrain NCDM models using the EDGES observations is based on the location of the minimum of the
absorption. Reference~\cite{Schneider:2018xba} imposed that it should
be located at a redshift higher than $z= 17.2$ and studied the
resulting bounds on a large set of NCDM models.  We show in the bottom
panels of Fig.~\ref{fig:xa20} the effect of the IDM scenario
considered here on the global sky-averaged 21~cm brightness
temperature obtained by means of Eq.~(\ref{eq:Tbdev}).

Let us focus first in the $f_*=0.01$ case (i.e. left
panel). Considering atomic cooling (dotted curves with $T_{\rm
  vir}^{\rm min}= 10^4$ K), it appears that for scattering cross
sections larger than $\sim 10^{-11} \, \sigma_T \times \left( {m_{\rm
    DM}}/{\rm GeV}\right)$ the absorption minimum takes place at
redshifts lower than $z= 17.2$. Such cross-sections should therefore be
regarded as disfavored.  Considering molecular cooling  softens this
constraint by $\sim$ one order of magnitude (see the continuous curves for $T_{\rm vir}^{\rm min}=
10^3$ K). Notice that we have considered a conservative X-ray
efficiency of $\zeta_X=5\times 10^{56} M_\odot^{-1}$ for all
curves. Considering lower values of $\zeta_X$ will give rise to a
later X-ray heating and thus a minimum of absorption located at lower
redshifts.

If instead, we consider a larger fraction of baryons converted into
stars, $f_*=0.03$, (see the bottom right panel), and $T_{\rm vir}^{\rm
  min}= 10^3$ $(10^4)$ K, the limit on the IDM interactions is
relaxed, excluding cross sections above $\sim 10^{-9}$ $( 10^{-10}) \,
\sigma_T \times \left( {m_{\rm DM}}/{\rm GeV}\right)$. This is due to
the fact that both the X-ray and the Lyman-$\alpha$ emission rates are
directly proportional to $f_*$. Increasing $f_*$ implies an earlier
Lyman-$\alpha$ coupling and X-ray heating periods, displacing the
minimum of the absorption in the 21cm signal to a larger redshift.

%%%%%%%%%%%%%%%%%%%%
\section{Discussion and Conclusions}
\label{sec:concl}
Interacting Dark Matter (IDM) models, in which dark matter is not made
out of purely cold, pressureless particles, are an interesting
alternative to the standard CDM paradigm and provide a possible avenue to
alleviate the so-called small-scale crisis of the $\Lambda$CDM. The
IDM could be scattering off light or massless degrees of freedom such as
photons or neutrinos. Here we have considered the case of dark matter
scattering on photons, characterized by the size of the scattering
rate over the DM mass, $\sidm/m_{\rm DM}$. The reason for this choice is driven
by the availability of a fitting function for the halo mass function
relevant for our study. Let us emphasize though that scatterings on
neutrinos are expected to give rise to a similar imprint on the 21 cm
signal.

Several studies have constrained IDM models based on their suppression
of clustering at small scales, exploiting galaxy power spectrum,
gravitational lensing, CMB, number of Milky Way dwarf galaxies and
Lyman-$\alpha$ forest observations, among others. Here we focus on the
imprint of IDM on the 21 cm signal arising from cosmic dawn.  Based on
a modified version of the {\tt 21cmFast} code, our simulations show
that IDM delays the formation of haloes capable of star formation, shifting the
timing of the 21 cm signal features compared to the standard CDM scenario.  A similar effect
has been reported in the case of other NCDM
models~\cite{Sitwell:2013fpa,Bose:2016hlz,Lopez-Honorez:2017csg,Escudero:2018thh}.
In this paper, we have considered two possible ways to test the IDM
properties against the 21 cm signal. First, following
Ref.~\cite{Lidz:2018fqo}, a significant Lyman-$\alpha$ coupling
between the gas and the spin temperature characterizing the 21 cm
signal should be present at $z=20$. Secondly, as argued in
Ref.~\cite{Schneider:2018xba}, the location of the absorption minimum
in the EDGES signal at $z=17.2$ implies that any scenario with
sufficiently enough delayed structure formation could be discarded.

We have first identified which are the most relevant astrophysical
parameters showing large degeneracies with the IDM scattering
cross-section over the mass $\sidm/m_{\rm DM}$. Namely, the fraction
of baryons into stars $f_*$, the threshold temperature for haloes to
host star-forming galaxies $T_{\rm vir}^{\rm min}$ and the X-ray
efficiency, $\zeta_X$, have been shown to interfere with the eventual 
extraction of a non-zero $\sidm$. Fortunately, the parameter $\zeta_X$ only plays a significant
role in extracting the location of the absorption minimum,  
and we have adopted the conservative value of $\zeta_X=5\times
10^{56} M_\odot^{-1}$ (corresponding to an integrated soft band X-ray
emissivity of $ L_{X<2{\rm keV}}/{\rm SFR}= 10^{40.2}$ erg/s /
($M_{\odot}$/yr)). We also considered a lower limit on the threshold
virial temperature of $T_{\rm vir}^{\rm min}= 10^3$~K (corresponding to
molecular cooling) as well as a constant value of $f_*<0.03$.

\begin{figure*}[t]
  \hspace*{-1cm}
\includegraphics[width=.5\textwidth]{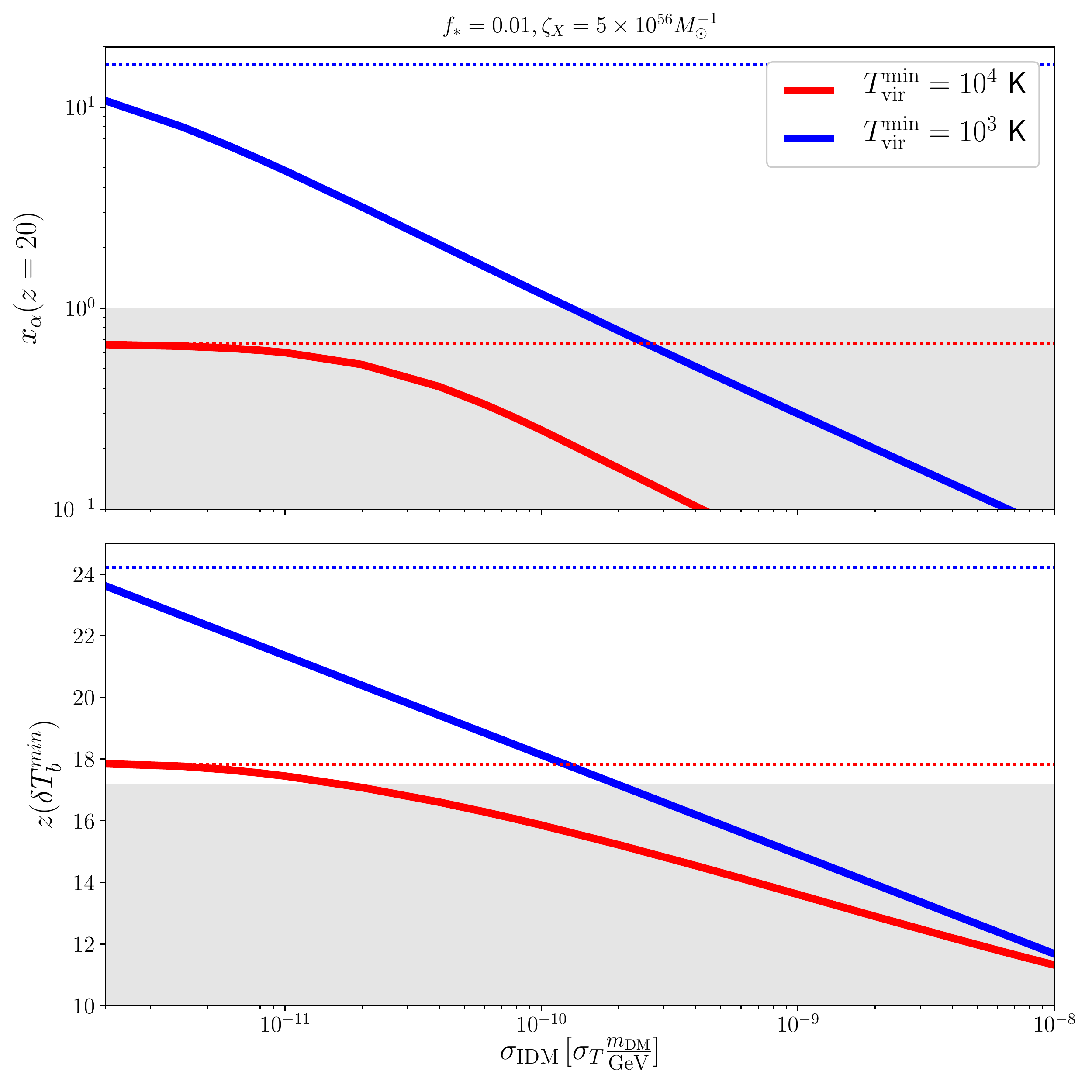}
\includegraphics[width=.5\textwidth]{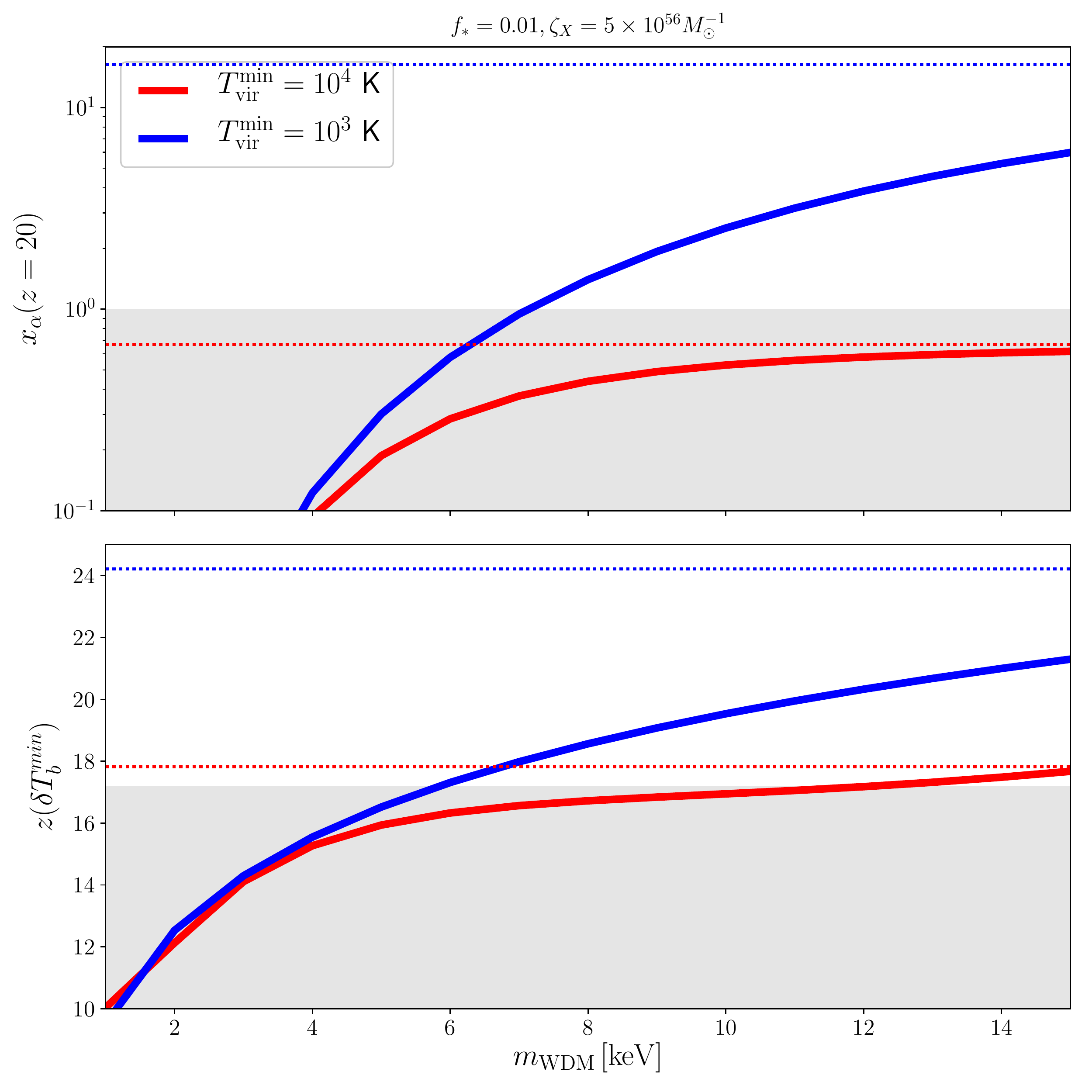}
\caption{Left, top (bottom) panel: Lyman-$\alpha$ coupling coefficient
  $x_\alpha$ at $z=20$ (redshift of the absorption minimum in the
  sky-averaged 21~cm brightness temperature) versus the scattering
  dark matter-photon cross-section for $T_{\rm vir}^{\rm min}=10^4$~K
  (atomic cooling) and for $T_{\rm vir}^{\rm min}=10^3$~K (molecular
  cooling). Right panels: the same but for WDM models, as a function
  of the WDM mass $m_{\rm WDM}$. IDM/WDM scenarios lying in the shaded
  regions are highly disfavoured by the two conditions exploited here,
  see text for details. We have fixed $f_*=0.01$ and $\zeta_X=5 \times
  10^{56} M_\odot^{-1}$. Dashed lines stand for the results in the CDM
  scenario.}
\label{fig:final}
\end{figure*}

Our results are summarized on the left panel of Fig.~\ref{fig:final}
in the case of $f_*=0.01$. The top panels shows the Lyman-$\alpha$
coupling coefficient $x_\alpha$ at $z=20$ for $T_{\rm vir}^{\rm
  min}=10^3$ ($10^4$)~K with the top blue (bottom red) curve as a
function of the IDM scattering cross-section. Notice that the
$x_{\alpha}(z=20)$ curves saturate to a maximum value at low enough
values of the scattering cross-section. This corresponds to the limiting $x_{\alpha}(z=20)$
value that one would get in the CDM scenario (indicated with the horizontal dotted lines). In addition, the shaded region refers to the
parameter space in which the condition $x_\alpha (z=20)>1$ can not be
satisfied and can therefore be considered as disfavored by the
condition given by Eq.~(\ref{eq:xa20}). As a result, for $f_*=0.01$ and
$\zeta_X=5 \times 10^{56} M_\odot^{-1}$, a value of $\sidm> 10^{-10}\,
\sigma_T \times \left( {m_{\rm DM}}/{\rm GeV}\right)$ fails to satisfy
the condition $x_\alpha (z=20)>1$ when considering molecular cooling
($T_{\rm vir}^{\rm min}=10^3$ K). This improves the previous bound
derived on such IDM model in Ref.~\cite{Escudero:2018thh}.  When
considering higher threshold temperatures for efficient cooling, the
bounds gets tighter, disfavouring even the canonical CDM scenario if
$T_{\rm vir}^{\rm min}=10^4$~K. All these limits would be relaxed if
the fraction of baryons converted into stars is larger, as it was
illustrated in Fig.~\ref{fig:xa20} for $f_*=0.03$. For instance, for
$T_{\rm vir}^{\rm min}=10^3$~K, the limit would be reduced by one
order of magnitude (i.e. $\sidm< 10^{-9}\, \sigma_T \times \left( {m_{\rm
    DM}}/{\rm GeV}\right)$).

In the bottom left panel of Fig.~\ref{fig:final}, we show the redshift
at which the sky-averaged 21~cm brightness temperature exhibits its
minimum of absorption at cosmic dawn, $z(\delta T_b^{min})$, as a
function of the IDM scattering cross-section. Again $z(\delta
T_b^{min})$ saturates for low enough scattering cross-section to a
value corresponding to the the CDM limit denoted with a dotted
horizontal line. The shaded area denotes the region in which this
minimum is located below $z=17.2$. More concretely, in order to derive
the curves shown in the bottom panels of Fig.~\ref{fig:final}, we have
fitted the brightness temperature curves obtained from our simulations
to the flattened gaussian shape that the EDGES collaboration uses to model
the 21 cm absorption profile~\cite{Bowman:2018yin}. This procedure
allows us to extract the central frequency for each possible value of
$\sidm/m_{\rm DM}$, $T_{\rm vir}^{\rm min}$ and $f_*$, and also to constrain the
model by imposing that this mean frequency $\nu_0$ of our flattened
gaussian fits should lie at frequencies below the lower $99\%$
confidence interval (including estimates of systematic uncertainties)
reported by the EDGES collaboration for the $\nu_0=78$ MHz parameter.
From Fig.~\ref{fig:final}, the position of the minimum of absorption
disfavours the region $\sidm> 10^{-10}\, \sigma_T \times \left(
{m_{\rm DM}}/{\rm GeV}\right)$ ($\sidm> 10^{-11}\, \sigma_T \times
\left( {m_{\rm DM}}/{\rm GeV}\right)$) for $T_{\rm vir}^{\rm
  min}=10^3$~K ($T_{\rm vir}^{\rm min}=10^4$~K).  As in the case of
the $x_\alpha (z=20)>1$ condition, these limits would be relaxed by
one order of magnitude in the case of a value of $f_*=0.03$. Also, for
the astrophysical parameters considered here, we see that the
constraints from the Lyman-$\alpha$ coupling condition are typically
tighter than those arising from the position of the minimum of absorption.

Summarizing, for $f_*\simeq 0.01, \zeta_X< 5 \times 10^{56} M_\odot^{-1}$ and
$T_{\rm vir}^{\rm min}>10^3$~K, the bounds on $\sidm/m_{\rm DM}$ derived in this
work imply an order of magnitude improvement over the most constraining
existing limits in the literature~\cite{Escudero:2018thh,Boehm:2014vja}. Larger values of
$f_*$ may compromise these upgraded limits. In order to ease the comparison to other studies on NCDM scenarios, we
provide the results obtained following the same methodology in a
thermal warm dark matter scenario involving light dark matter
particles with a mass $m_{\rm{WDM}}$ of a few keV. The prescription
considered here to describe the suppression of the halo mass function
at small halo masses is given in the Appendix~\ref{sec:hmf}. We show the
obtained dependence of $x_\alpha (z=20)$ and $z(\delta T_b^{min})$ in
the right panel of Fig.~\ref{fig:final}.  From this figure with
$f_*=0.01$ and $\zeta_X= 5 \times 10^{56} M_\odot^{-1}$, we can infer a
lower limit in the WDM mass around $m_{\rm WDM}>6$~keV ($m_{\rm
  WDM}>12$~keV) if $T_{\rm vir}^{\rm min}=10^3$~K ($T_{\rm vir}^{\rm
  min}=10^4$~K). These tight limits are similar to those derived in
Refs.~\cite{Schneider:2018xba,Lidz:2018fqo} for slightly different WDM
implementations and astrophysical parameters.

\appendix
\section{Halo mass functions}
\label{sec:hmf}
The halo mass function, that counts the number of haloes per unit halo mass and volume at a given redshift, can be written as~\cite{Press:1973iz}
\begin{equation}
  \frac{dn}{dM}=-\frac12 \, \frac{\rho_m}{M^2} \, f(\nu) \, \frac{d\ln \sigma^2}{d\ln M} ~,
\label{eq:dndM}
\end{equation}
where $n$ is the halo number density, $\rho_m = \Omega_m \, \rho_c$ is
the average matter density in the Universe at $z = 0$ and $\sigma^2 =
\sigma^2(M,z)$ is the variance of density perturbations, which is a
function of the halo mass $M$ and redshift. The first-crossing
distribution, $f (\nu)$, is expected to be a universal function of
$\nu \equiv \delta_c^2/\sigma^2(M,z)$, with $\delta_c = 1.686$, the
linearly extrapolated density for collapse at $z = 0$. The Sheth-Tormen (ST) first-crossing
distribution reads as~\cite{Sheth:1999mn, Sheth:1999su, Sheth:2001dp}:
\begin{eqnarray}
f(\nu) = A \, \sqrt{\frac{2 \, q \, \nu}{\pi}} \left(1 + (q \, \nu)^{-p}\right) \, e^{-q \, \nu/2} ~, 
\label{eq:ST}
\end{eqnarray}
with $p = 0.3$, $q = 0.707$ and $A = 0.3222$. For the standard CDM
scenario, we consider this first-crossing distribution.  The variance
at $z = 0$, $\sigma(M) \equiv \sigma(M, z = 0)$, at a given scale $R$
can be expressed as
\begin{equation}
\sigma_X^2(M(R)) = \int \frac{d^3k}{(2\pi)^3} \, P_X(k) \, W^2(kR) ~, 
\label{eq:sig2}
\end{equation}
where $P_X(k)$ is the linear power spectrum at $z = 0$ for a given $X
= \{\textrm{CDM, IDM or WDM}\}$ cosmology and $W$ is the Fourier
transform of a filter function that we consider to be a top-hat (TH)
function in real space (see
e.g.~\cite{Murgia:2017lwo,Schneider:2018xba} for the possibility of
using a sharp-$k$ window function for NCDM). The redshift dependence is
driven by the linear growth function, $D(z)$ normalized to 1 at $z =
0$, so that the root-mean-square (rms) density fluctuation is
$\sigma(M, z) = \sigma(M, z = 0) \, D(z)$.

The transfer function $T_X$ for a NCDM scenario $X$ is defined
as
\begin{equation}
P_{X}(k) = P_{\rm{CDM}}(k) \, T^2_{X}(k) ~,
\label{eq:pwdm}
\end{equation}
where $P_{\rm{CDM}}(k)$ is the linear power spectrum in
$\Lambda$CDM. Here we use the prescription of Refs.~\cite{Bode:2000gq,
  Boehm:2001hm}:
\begin{equation}
T_{X}(k) = \left(1+ (\alpha_{X} k)^{2\mu}\right)^{-5/\mu} ~,
\label{eq:twdm}
\end{equation}
where $\mu = 1.2$ is a dimensionless exponent and
$\alpha_X$ is a breaking scale. The latter takes the form:
\begin{widetext}
\begin{eqnarray}
\alpha_{\rm{WDM}} &=& 0.048 \left(\frac{{\rm keV}}{m_{\rm{WDM}}}\right)^{1.15} \left(\frac{\Omega_{\rm{WDM}}} {0.4}\right)^{0.15}\left(\frac{h}{0.65}\right)^{1.3} \, {\rm Mpc}/h ~,
\label{eq:alphWDM}\\
\alpha_{\rm{\gamma DM}}&=& 0.073 \, \left[10^8 \, \left(\frac{\sidm}{\sigma_T}\right) \, \left(\frac{{\rm GeV}}{m_{\rm DM}}\right)\right]^{0.48} \left(\frac{\Omega_{\rm{WDM}}} {0.4}\right)^{0.15}\left(\frac{h}{0.65}\right)^{1.3} \, {\rm Mpc}/h ~\,.
\label{eq:alphIDM}
\end{eqnarray}
\end{widetext}
for WDM scenarios~\cite{Schneider:2011yu, Schewtschenko:2014fca} and
IDM involving dark matter-photon scattering~\cite{Boehm:2001hm}
respectively.  For DM-neutrino interactions, described with the same
parametrization of the transfer function, one gets a breaking scale
$\alpha_{\rm{\nu DM}}\simeq 0.8 \times \alpha_{\rm{\gamma
    DM}}$ for a fixed value of the scattering cross section~\cite{Schewtschenko:2014fca}.

 The halo mass function defined as in Eq.~(\ref{eq:dndM}) is well
 suited for CDM but it needs to be adapted  for the NCDM case. In order to
 match the results from N-body simulations, the WDM halo mass function
 can be expressed as~\cite{Schneider:2011yu}
\begin{equation}
\frac{dn^{\rm WDM}}{dM} = \left(1 + \frac{M_{\rm hm}}{b \, M}\right)^{a}  \frac{dn^{\rm ST, \, WDM}}{dM} ~,
\label{eq:WDM}
\end{equation}
where an additional mass-dependent correction to the standard ST
formalism appears. We use $a=-0.6$ and $b=0.5$, as obtained in~\cite{Schewtschenko:2014fca, Moline:2016fdo}.
 The function $\frac{dn^{\rm ST, \, WDM}}{dM}$ in Eq.~(\ref{eq:WDM})
 refers to the halo mass function obtained with a ST first-crossing
 distribution, as defined in Eq.~(\ref{eq:ST}), and a linear matter
 power spectrum corresponding to the WDM case.  In order to describe
 the suppression in the linear regime, one can consider the half-mode
 mass $M_{\rm hm}$, defined as the mass scale for which $T_{X}/T_{\rm
   CDM} = 1/2$ (i.e. $P_{X}/P_{\rm CDM} = 1/4$). Using the general
 fit to the transfer function, Eq.~(\ref{eq:twdm}), the half-mode mass
 $M_{\rm hm}$ can be easily derived as
\begin{eqnarray}
M_{\rm hm} & \equiv & \frac{4 \, \pi}{3} \, \rho_m \, \left(\pi  \alpha_X \right)^3 \left(2^{\mu/5} - 1 \right)^{-3/(2 \, \mu)} ~.
\label{eq:Mhm}
\end{eqnarray}
For what concerns the IDM models, the number of low-mass structures
appears to be larger than in WDM
scenarios~\cite{Schewtschenko:2014fca} (see also
\cite{Vogelsberger:2015gpr}). In order to reproduce the IDM results
for masses below the half-mode mass, an extra mass-dependent
correction must be introduced to the halo mass
function~\cite{Moline:2016fdo}:
\begin{equation}
\frac{dn^{\rm IDM}}{dM} =\left(1 + \frac{M_{\rm hm}}{b \, M}\right)^{a} \left(1+\frac{M_{\rm hm}}{g \, M}\right)^{c} \, \frac{dn^{\rm ST,\, CDM}}{dM} ~,
\label{eq:IDM}
\end{equation}
with $a = -1$, $b = 0.33$, $g = 1$, $c = 0.6$ and $\frac{dn^{\rm ST,
    \, CDM}}{dM}$ refers to the standard ST first-crossing
distribution as defined in Eq.~(\ref{eq:ST}) with the CDM
linear power spectrum for the variance of density perturbations.

\section{Lyman-$\alpha$ emissivity}
\label{sec:appb}

The Lyman-$\alpha$ flux from direct stellar emission of UV photons,
$J_{\alpha\star}$, is given by the sum over the Lyman-$n$
levels which can lead to a $2p \rightarrow 1s$ transition through a
decaying cascade. Due to the optical thickness of the IGM, photons
which redshift to a Lyman resonance are absorbed by the medium. A
photon which reaches a Lyman-$n$ resonance at redshift $z$ has to be
emitted at a redshift below $1+z_{max,n} =
\frac{1-(n+1)^{-2}}{1-n^{-2}}(1+z)$. If $f_{rec}(n)$ is the recycled
fraction of the level $n$, \textit{i.e.}, the probability of
generating a Lyman-$\alpha$ photon from the $n$ level
\cite{Pritchard:2005an}, the total Lyman-$\alpha$ flux can be written as
\begin{equation}
J_{\alpha\star}=\frac{c(1+z)^2}{4\pi} \sum_{n=2}^{n_{max}} f_{rec}(n) \int_z^{z_{max,n}} dz'\frac{\epsilon_{\alpha}(\nu'_n,z')}{H(z')}~,
\end{equation}
where the emission frequency is $\nu'_n= \nu_n\frac{1+z'}{1+z}$, being $\nu_n=\nu_{LL}(1-n^{-2})$ and $\nu_{LL}$ the Lyman limit frequency. The comoving emissivity $\epsilon_{\alpha}$ can be written as proportional to the star formation rate $\dot{\rho}_{\star}$:
\begin{equation}
\epsilon_{\alpha}(\nu,z) = \varepsilon(\nu) \frac{\dot{\rho}_{\star}(z)}{\mu m_p} = \varepsilon(\nu) f_* \bar{n}_{b,0}  \dot{f}_{coll}(z),
\end{equation}
where $\bar{n}_{b,0}$ is the comoving number density of
baryons. Assuming that only Population-II stars contribute to this
emissivity, the spectral distribution $\varepsilon(\nu)$ is given by a
separate power law between each Lyman-$n$ and Lyman-$n+1$ levels:
\begin{equation}
\varepsilon(\nu)=N_n\frac{(\beta_n+1)\nu_{\alpha}^{\beta_n}}{(\nu_{n+1}^{\beta_n+1}-\nu_{n}^{\beta_n+1})} \left( \frac{\nu}{\nu_{\alpha}}\right)^{\beta_n}
\end{equation}
for $\nu_n \leq \nu \leq \nu_{n+1}$, with $N_n$ the number of photons emitted between the $n$ and $n+1$ resonances and $\beta_n$ the spectral index \cite{Barkana:2004vb}. The function above is normalized as
$\int_{\nu_n}^{\nu_{n+1}} d\nu \varepsilon(\nu) = N_n$, with
$\int_{\nu_{\alpha}}^{\nu_{LL}} d\nu \varepsilon(\nu) = \sum_n N_n
\simeq 9690$ the total number of photons emitted between the
Lyman-$\alpha$ and the Lyman limit. Although we keep this
normalization as constant through our analysis, 
notice that changes in the number of photons which contribute to the 
Lyman-$\alpha$ flux could have a deep impact in the 21cm signal \cite{Witte:2018itc}.

\begin{comment}

\begin{equation}
J_{\alpha}=\frac{c(1+z)^2}{4\pi}\int_z^{\infty} dz' \frac{dt'}{dz'} (1+z')\hat{\epsilon}_{\alpha}(z')
\end{equation}

\begin{equation}
\hat{\epsilon}_{\alpha}(z') = \sum_{n=2}^{n_{max}} f_{rec}(n)\varepsilon(\nu'_n) \theta(z_{max,n} - z') \frac{\dot{\rho}_{\star}}{\mu m_p} 
\end{equation}

\begin{equation}
\varepsilon(\nu)=\varepsilon_n(\nu) \; \; \; for \; \; \; \nu_n \leq \nu \leq \nu_{n+1}~,
\end{equation}
with
\begin{equation}
\varepsilon_n(\nu)=N_n\frac{(\beta_n+1)\nu_{\alpha}^{\beta_n}}{(\nu_{n+1}^{\beta_n+1}-\nu_{n}^{\beta_n+1})} \left( \frac{\nu}{\nu_{\alpha}}\right)^{\beta_n}.
\end{equation}
The function above is normalized as
\begin{equation}
\int_{\nu_n}^{\nu_{n+1}} d\nu \varepsilon_n(\nu) = N_n, \; \; \; \int_{\nu_{\alpha}}^{\nu_{LL}} d\nu \varepsilon(\nu) = \sum_n N_n \simeq 9690.
\end{equation}

\begin{widetext}
\begin{equation}
J_{\alpha}=\frac{c(1+z)^2f_{\star}n_{b,0}}{4\pi} \sum_{n=2}^{n_{max}} f_{rec}(n) \int_z^{z_{max,n}} dz'(1+z')(1+\delta)\frac{df_{coll}}{dz'}\varepsilon(\nu'_n)
\end{equation}
\end{widetext}

\end{comment}

\medskip

\acknowledgments

LLH is supported by the FNRS, by the Strategic Research Program {\it
  High Energy Physics} and by the Research Council of the Vrije
Universiteit Brussel. OM and PVD are supported by PROMETEO
II/2014/050, by the Spanish MINECO Grants FPA2014–57816-P,
FPA2017-85985-P and SEV-2014-0398 and by the European Union’s Horizon
2020 research and innovation program under the Marie Sklodowska-Curie
grant agreements No. 690575 and 674896.

\bibliography{biblio.bib}
\bibliographystyle{JHEP}

\end{document}